\documentclass[english,aps, prl, showpacs, preprint]{revtex4}
\usepackage[T1]{fontenc}
\usepackage[latin1]{inputenc}
\usepackage{array}
\usepackage{graphicx}

\makeatletter

\providecommand{\tabularnewline}{\\}

\usepackage{array}

\makeatletter

\usepackage{array}


\makeatother

\usepackage{babel}
\makeatother
\begin{document}

\title{Chaotic dynamics of spin valve oscillators}

\author{Z. Yang}

\author{S. Zhang}

\affiliation{Department of Physics and Astronomy, University of Missouri-Columbia,
Columbia, Missouri 65211}

\author{Y. Charles Li}

\affiliation{Department of Mathematics, University of Missouri-Columbia, Columbia,
Missouri 65211}

\begin{abstract}
Recent experimental and theoretical studies on the magnetization dynamics
driven by an electric current have uncovered a number of unprecedented
rich dynamic phenomena. We predict an intrinsic chaotic dynamics that
has not been previously anticipated. We explicitly show that the transition
to chaotic dynamics occurs through a series of period doubling bifurcations.
In chaotic regime, two dramatically different power spectra, one with
a well-defined peak and the other with a broadly distributed noise,
are identified and explained. 
\end{abstract}

\date{\today}


\maketitle
Among many well-studied non-linear oscillators driven by external
forces, only a handful oscillators have technological applications
\cite{Pikovsky,Rabinovich}. The recently discovered current-driven
magnetization oscillators with tunable microwave frequencies in spin
valves are very desirable for magnetic storage devices and for telecommunications.
Up until now, theoretical and experimental studies of the oscillator
have been carried out only for the simplest cases where the dynamics
of the oscillator is either in a self-sustained steady-state precessional
motion (limit cycle) \cite{Slonczeweski,Sun,Kiselev,Krivorotov,Li,Bertotti}
or in synchronization with other oscillator(s) \cite{Rippard,Kaka,Mancoff,Grollier,Slavin}.
Since the equation that governs the oscillator is highly non-linear,
it would be fundamentally interesting to map out the full dynamics
for experimentally relevant parameters. In particular, a thorough
study of chaotic dynamics will elucidate how the current-driven oscillator
responds to an external perturbation.

Previous study of the current-driven magnetization chaos or noise
was based on the micromagnetic simulation where the magnetization
is not uniform due to strong magnetostatic interaction at the edge
of the sample \cite{Lee,Berkov}. These chaotic dynamics highly depend
on the shape and size of the sample, and thus it is not an intrinsic
property of the current-driven oscillator. Here we consider a single-domain
current-driven spin valve oscillator so that the undesired complication
of the spatial variation of the magnetization is eliminated. The intrinsic
dynamic property of the spin valve oscillator is investigated in the
presense of an external periodic perturbation, for example, an AC
current. By utilizing the Poincar\'{e} map \cite{Alligood}, we have
found the route to chaos to be via period doubling bifurcations. Positive
Lyapunov exponents\cite{Alligood} and Sharkovskii ordering\cite{Ott}
are observed as the evidence of chaos. Furthermore, we show two dramatically
different power spectra in chaotic regions, one with a well-defined
peak and the other with broadly distributed noise.

The modeled spin valve consists of a pinned layer whose magnetization
is fixed along the positive $x$-axis and a single-domain free layer
whose magnetization vector ${\bf m}$ is the subject of our calculation.
The free layer experiences an effective field ${\bf H}_{eff}$ made
of an external field, an anisotropy and a demagnetization field perpendicular
to the layer ($z$-axis). We choose the direction of the magnetic
field along the in-plane easy axis \cite{footnotes}. The dynamics
of the magnetization on the free layer is determined by the modified
Landau-Lifshitz-Gilbert (LLG) equation. \cite{Slonczeweski} \begin{equation}
\frac{\partial{\bf m}}{\partial t}=-\gamma{\bf m}\times\mathbf{H}_{eff}+\alpha{\bf m}\times\frac{\partial{\bf m}}{\partial t}+a_{j}\gamma{\bf m}\times({\bf m}\times{\bf e}_{x})\label{eq:llg}\end{equation}
 where $a_{j}$ is the amplitude of the spin torque which has the
unit of the magnetic field (Oe). Since the magnitude of the magnetization
$|{\bf m}|=1$ is a constant, there are only two independent variables,
thus, chaotic dynamics are excluded for a constant $a_{j}$. In fact,
the solutions of Eq.~(\ref{eq:llg}) have already been obtained \cite{Bertotti,Serpicoa}.
The time-dependent solution is a limit cycle which can be analytically
determined via the construction of a Melnikov integral (MI) \cite{Bertotti}.
It has been found that the stable limit cycle with a well defined
frequency (which we call the natural frequency $\omega_{0}$) exists
only for the current density larger than a critical current density.
Two distinct limit cycles have been identified: an out-of-plane orbit
and a nearly in-plane orbit around ${\bf m}={\bf e}_{x}$.

To explicitly reveal the dynamics of the above spin valve oscillator
under an external perturbation, we now consider a time-dependent current
which adds an additional term in Eq.~(\ref{eq:llg}), $a_{ac}\cos(\omega t)\gamma{\bf m}\times({\bf m}\times{\bf e}_{x})$.
The simplest dynamic phase would be the synchronization, i.e., the
spin valve oscillator is forced to oscillate in phase with the external
frequency ${\bf \omega}$ as long as $\omega_{0}$ is sufficiently
close to $\omega$. Indeed, the recent experiment \cite{Rippard}
and theory \cite{Grollier,Slavin} have clearly demonstrated the synchronization.
The much richer dynamics, however, would be chaotic dynamics \cite{Li_Chaos}
shown below.

\begin{figure}
\includegraphics[bb=200bp 0bp 580bp 390bp,scale=0.3]{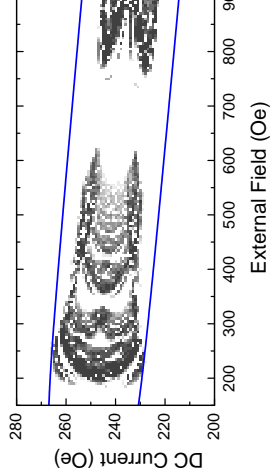}

\caption{Chaotic/non-chaotic phase diagram. Two lines are the boundary of
chaos determined by Melnikov Integral. Chaos is possible only within
the region bounded by these two lines. The dark region represents
a positive value of the largest Lyapunov exponent, i.e., chaos. The
parameters are chosen similar to a permalloy film: damping constant
$\alpha=0.02$, anisotropy constant $H_{K}=0$, demagnetization field
$4\pi M_{s}=8400(Oe)$ and $\gamma=1.7\times10^{-7}(Oe^{-1}s^{-1})$.
The amplitude and the frequency of the ac current are $a_{ac}=20(Oe)$
and $\omega=15(GHz)$. \label{fig:chaosphase}}
\end{figure}

An analytical prediction of chaotic dynamics can be made using a MI
along the separatrix orbit. The simple zeros of the MI indicate the
occurrence of chaotic dynamics \cite{Li_Chaos}. In Ref.~\cite{Li_Chaos},
the MI was carried out for a fixed magnetic field. Here we map out
the phase diagram in a parameter space of both DC current and magnetic
field. Implementing the MI approach, we obtain a boundary loop which
encloses all the possible simple zeros in the parameter space. In
Fig.~\ref{fig:chaosphase}, the phase diagram is plotted. Two solid
lines are a portion of the complete boundary loop; if we extend the
parameter space to a much larger range, the two lines will meet to
form a closed area. Since there exists an arbitrary phase in calculating
the MI, the exact solutions are not available. Therefore, a more rigorous
numerical test is required to identify the true chaos bounded by the
boundary lines. We show below that only the dark region in Fig.~\ref{fig:chaosphase}
is chaos.

\begin{figure}
\includegraphics[bb=200bp 0bp 580bp 580bp,scale=0.3]{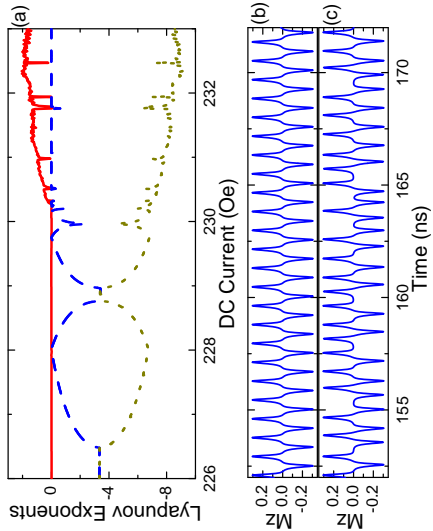}

\caption{(a) Lyapunov spectra for $226(Oe)<a_{j}<233(Oe)$; (b) the regular
non-chaotic trajectory for $a_{j}=227.5(Oe)$ ; (c) chaotic trajectory
for $a_{j}=232.5(Oe)$. The magnetic field is 200 Oe. All parameters
are same as in Fig.~1.}

\label{fig:lyapunov} 
\end{figure}

Numerically, chaos is often indicated by the positive Lyapunov exponents.
The Lyapunov exponent measures the exponential increase (or decrease)
of an initial distance from two close trajectories, \[
\lambda_{i}=\lim_{t\rightarrow\infty}\frac{1}{t}\ln\frac{\left\Vert \delta m_{t}^{i}\right\Vert }{\left\Vert \delta m_{0}^{i}\right\Vert }\]
 where $\lambda_{i}$ is the $i$th Lyapunov exponent and $\left\Vert \delta m_{t}^{i}\right\Vert $
is the distance between the trajectories of the $i$th orthogonal
axis at time $t$. When $\lambda_{i}>0$, the distance between two
arbitrarily close points will increase exponentially along the $i$-axis
at a large $t$ \cite{Gonzalez-Miranda}. For a bounded dynamic system,
any positive Lyapunov exponent indicates chaos. We numerically computed
the Lyapunov exponents and Fig.~\ref{fig:lyapunov}(a) shows an example
of the Lyapunov spectra, where the largest exponent is shown in red
solid line and the other two are in dashed and dotted lines. When
$a_{j}$ is smaller than the critical value $230.03(Oe)$, all three
Lyapunov exponents are non-positive and thus the dynamics are non-chaotic,
see Fig.~\ref{fig:lyapunov}(b). When $a_{j}$ becomes larger than
$230.03(Oe)$, the largest Lyapunov exponents become positive, and
thus chaos appears {[}Fig.~\ref{fig:lyapunov}(c)]. By sweeping the
parameters of the DC current and the external field, we have mapped
out the parameters that give arise at least one positive Lyapunov
exponent, shown as the dark area in Fig.~\ref{fig:chaosphase}.

To understand how chaos develops, we applied the Poincar\'{e} map
to this system. In spin valve oscillators, it is difficult to choose
a simple Poincar\'{e} section because a simple section can hardly
include both out-of-plane and in-plane orbits simultaneously. To avoid
this difficulty, we record the trajectory points every time a local
minimum of $m_{x}$ is reached. If we define the time intervals between
two successive minima as a period, this is essentially a Poincar\'{e}
period map except that the period is not a constant.

\begin{figure}
\includegraphics[bb=200bp 0bp 582bp 582bp,scale=0.3]{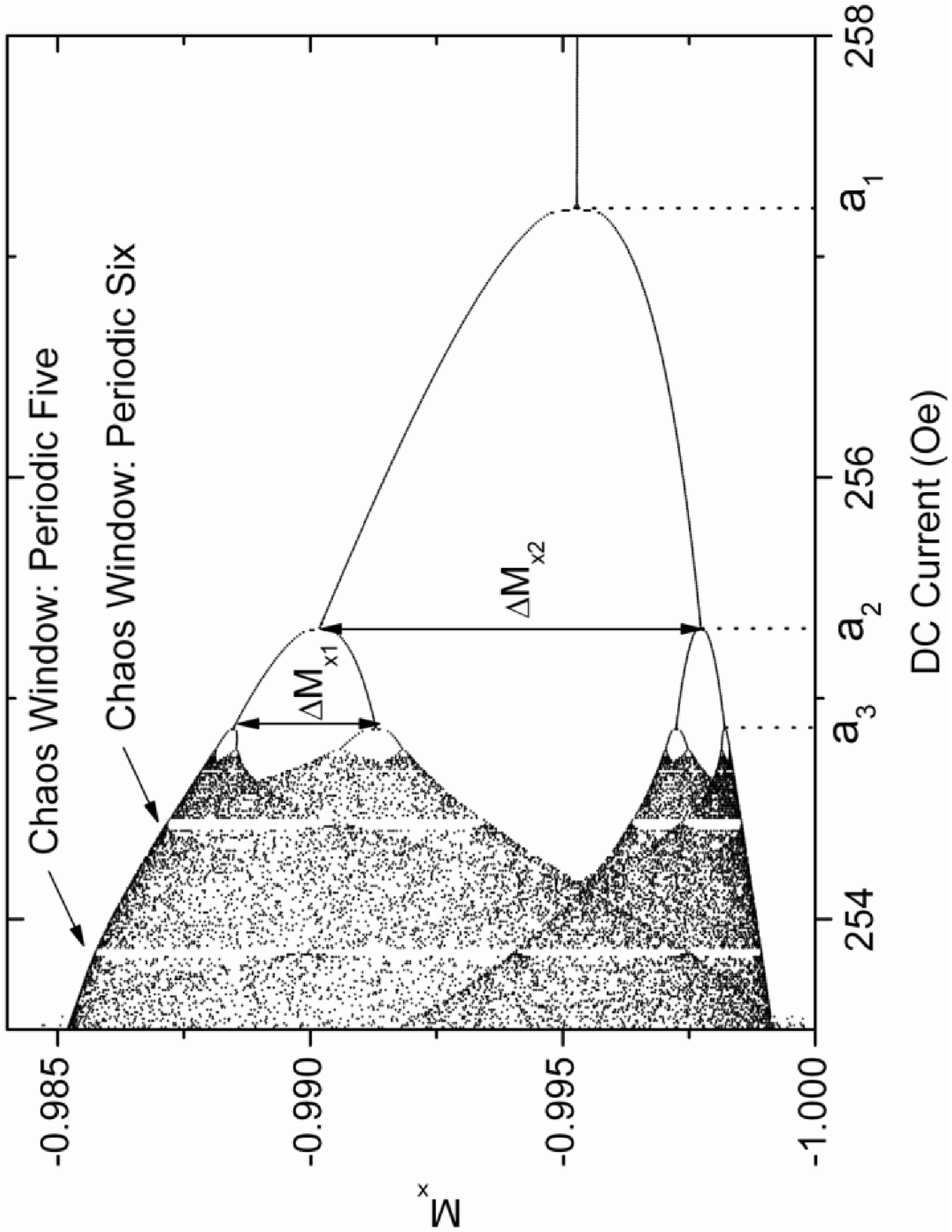}

\caption{Period doubling bifurcation cascade. Dashed lines indicate the critical
values of the DC current where the bifurcation occurs. All parameters
are the same as in Fig.~2.}

\label{fig:cascade} 
\end{figure}

By using the minimum $m_{x}$ map, we are able to see the development
of chaotic dynamics when one varies the parameters. For limit cycles
where the parameters are within the non-chaotic white regions shown
in Fig.~1, the magnetization follows a unique trajectory and the
map point is just one unique point. When the parameters, e.g., the
currents, are close to the boundaries between the white and dark regions
of Fig.~1, two map points are seen; this is identified as the period
doubling bifurcation. In this case, the magnetization orbit will return
to the original orbit after two periods. If we choose the current
even closer to the boundaries, a period-four orbit and, in general,
a period-$2^{n}$ orbit, appear. When $n\rightarrow\infty$, the magnetization
dynamics become chaotic, i.e., the appearance of the dark regions
in Fig.~1. In Fig.~\ref{fig:cascade}, we have shown an example
of the bifurcation diagram where the transition from the synchronization
to the bifurcation and to chaos is clearly demonstrated. Period-five
and period-six chaos windows are observed between 252 to 253 Oe in
the cascade. According to Sharkovskii ordering theorem, these observations
imply the existence of chaos \cite{Ott}.

The period doubling bifurcation cascade shown in Fig.~\ref{fig:cascade}
is in fact similar to those in other nonlinear systems \cite{Alligood}.
To see if our bifurcation structure belongs to the same universal
quadratic class, we evaluate the Feigenbaum ratio\cite{Feigenbaum}
which is defined as \[
\delta_{i}=\frac{a_{i}-a_{i+1}}{a_{i+1}-a_{i+2}}\]
 where $a_{i}$ is the value of the parameter at the $i$th bifurcation
point. In Table~\ref{tab:Feigenbaum-ratio-measurement}, We list
an example of measurement of Feigenbaum ratios for our system. There
is a second Feigenbaum ratio that denotes the rescaling ratio of the
bifurcation forks. For example, from first two bifurcation in Fig.\,\ref{fig:cascade},
we measured $|\alpha|=|\Delta M_{x1}/\Delta M_{x2}|=2.68$. Analytically,
the Feigenbaum ratios are approaching to the universal constants 4.6692\ldots{}
for $\delta$ and 2.5029\ldots{} for $|\alpha|$ \cite{Feigenbaum},
for a quadratic non-linear map\cite{VanDerWeele}. We find that the
measured ratios agree with the universal Feigenbaum numbers exceedingly
well. This indicates that our system indeed belongs to a quadratic
nonlinear class.

\begin{table}

\caption{Feigenbaum ratio measurement. The second column is derived from Fig.~3;
$a_{i}$ is a critical value at the $i$th bifurcation point. }

\begin{tabular}{|>{\centering}p{15mm}|>{\centering}p{15mm}|>{\centering}p{15mm}|>{\centering}p{15mm}|}
\hline 
$i$&
$a_{i}$ &
$a_{i}-a_{i+1}$&
$\delta_{i}$\tabularnewline
\hline 
1&
257.2500 &
-1.9300&
4.289\tabularnewline
2&
255.3200&
-0.4500&
4.356\tabularnewline
3&
254.8700&
-0.1033&
4.472\tabularnewline
4&
254.7667 &
-0.0231&
4.529\tabularnewline
5&
254.7436&
-0.0051&
4.636\tabularnewline
6&
254.7385&
-0.0011&
\tabularnewline
7&
254.7374 &
&
\tabularnewline
\hline
\end{tabular}\label{tab:Feigenbaum-ratio-measurement} 
\end{table}

By summing all Lyapunov components,we have verified that the volume
contraction ($\lambda_{v}=\lambda_{1}+\lambda_{2}+\lambda_{3}$) is
always negative even though the Lyapunov exponent can be positive;
the negative volume contraction indicates the dissipative nature of
the spin valve oscillator \cite{Alligood,Gonzalez-Miranda}. %the volume enclosed by the initial distance sphere (see Ref.~\cite{Iaxis}) exponentially contracts at the large $t$
%limit. Thus, the spin valve oscillator is indeed dissipative \cite{Alligood}.

\begin{figure}
\includegraphics[bb=200bp 0bp 582bp 582bp,scale=0.33]{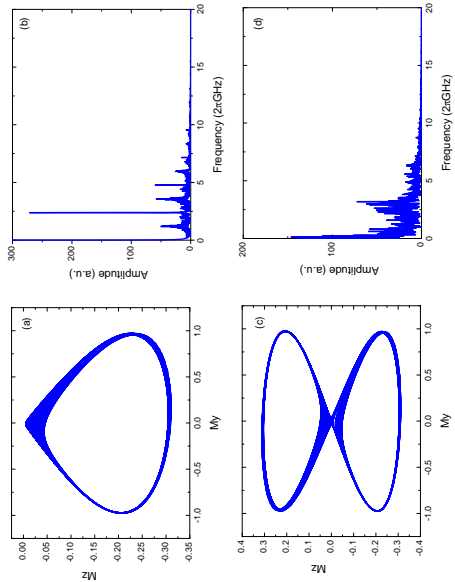}

\caption{Trajectories and their corresponding $M_{z}$-component power spectra
of two types of chaos. Upper panels: $a_{j}=262.5(Oe)$; the largest
Lyapunov exponent is 0.86. Lower panels: $a_{j}=261.0(Oe$); the largest
Lyapunov exponent is 1.56. All parameters are same as in Fig.~2.
\label{fig:powerspectra}}
\end{figure}

A highly interesting feature is the two distinct magnetization trajectories
during the transition to chaos. In the first case, the bifurcation
occurs only at the out-of-plane orbits shown in Fig.\,\ref{fig:powerspectra}(a).
Although the bifurcation on this single orbit also leads to chaos
because the largest calculated Lyapunov exponent is positive, the
power spectrum displays a well defined peak. The peak position corresponds
to the inverse of the average time for the magnetization to complete
one loop (since the loop never closes, we define a one-loop when the
trajectory returns to the point nearest to the starting point of the
loop). Thus, the presence of the narrow peak indicates a quasi-periodic
motion of the magnetization in chaotic dynamics. It would be erroneous
if one automatically assumes the dynamics is synchronization when
the experimental power spectrum is highly peaked. Synchronization
refers to the phase-locking between the external and natural frequencies
but the positive Lyapunov exponent excludes the possibility of the
phase-locking. The second chaotic motion involves both out-of-plane
and in-plane orbits, as seen in Fig.~\ref{fig:powerspectra}(c).
In this case, the power spectra display typical noise, i.e., broadly
distributed spectra. The magnetization jumps between out-of-plane
and in-plane orbits are completely random; it is an intrinsic stochastic
process driven by a deterministic external perturbation. This stochastic
jumping leads to a much stronger noise in the power spectra. On the
other hand, it is necessary that trajectories are within the vicinity
of the separatrix of two different orbits so that the stochastic magnetization
jump can take place under small perturbations. Since the separatrix
has an infinite period, the trajectory close to it must have a nearly
zero frequency as seen in Fig.\,\ref{fig:powerspectra}(d).

The synchronization-like chaotic power spectra imply that quasi-periodicity
can be observed in chaos if the trajectory is not near the separatrix.
It is the stochastic jump of the trajectory in the vicinity of the
separatrix that reduces the periodicity. The reduction of the periodicity
due to stochastic jumps could also occur when the thermal fluctuations
exist because the thermal fluctuations are stochastic, known as a
Wiener process. In general, there exists a bifurcation gap \cite{Crutchfield}
where the thermal fluctuations limit the observation of $2^{n}$ orbits
in the bifurcation diagram to a finite number $n<n_{0}$. The stronger
the fluctuations are, the smaller $n_{0}$ is. Consequently, the thermal
fluctuations make the Lyapunov spectra smoother in Fig. 2 and chaos
windows invisible in Fig. 3. We emphasize that although the thermal
fluctuations lead to noise in the power spectra and the possible random
jumps between two orbits, it is not associated with the bifurcation
and thus there are no universal Feigenbaum numbers. Experimentally,
one can distinguish chaotic dynamics generated by the deterministic
perturbation (AC current) and by the thermal fluctuations. For example,
when the parameter, e.g. DC current in our study, is varied, the current
driven power spectrum should evolve from non-chaotic dynamics to chaos
and back to non-chaos.

In conclusion, by using both analytical and numerical approaches,
we have shown the route to chaos and that the oscillator belongs to
the class of dissipative quadratic non-linear systems. Our findings
demonstrate that the current-driven magnetization dynamics is much
richer than previously studied steady-state motion and synchronization.
The magnetization oscillator whose dynamics can be measured by experiments
provides a model system to verify chaos theories in a general non-linear
system.

This work was supported by DOE(DE-FG02-06ER46307).


\begin{thebibliography}{10}
\bibitem{Pikovsky}A. Pikovsky, M. Rosenblum, and J. Kurths, \textit{Synchronization:
A Universal Concept in Nonlinear Sciences} (Cambridge University Press,
Cambridge, 2001).

\bibitem{Rabinovich}M. I. Rabinovich and D. I. Trubetskov, \textit{Oscillations
and Waves in Linear and Nonlinear Systems} (Kluwer, Dordrecht, 1989).

\bibitem{Slonczeweski}J. C. Slonczewski, J. Magn. Magn. Mater. \textbf{159},
L1(1996); \textbf{195}, 261 (1999).

\bibitem{Sun}J. Z. Sun, Phys. Rev. B \textbf{62}, 570 (2000).

\bibitem{Bertotti}G. Bertotti, C. Serpico, I. D. Mayergoyz, A. Magni,
M. d'Aquino, and R. Bonin, Phys. Rev. Lett. \textbf{94}, 127206 (2005).

\bibitem{Kiselev}S. I. Kiselev, J. C. Sankey, I. N. Krivorotov, N.
C. Emley, R. J. Schoelkopf, R. A. Buhrman, and D. C. Ralph, Nature
(London) \textbf{425}, 380(2003).

\bibitem{Krivorotov}I. N. Krivorotov, N. C. Emley, J. C. Sankey,
S. I. Kiselev, D. C. Ralph, and R. A. Buhrman, Science \textbf{307},
228(2005).

\bibitem{Li}Z. Li and S. Zhang, Phys. Rev. B \textbf{68}, 024404
(2003).

\bibitem{Rippard}W. H. Rippard, M. R. Pufall, S. Kaka, T. J. Silva,
S. E. Russek, and J. A. Katine, Phys. Rev. Lett. \textbf{95}, 067203
(2005).

\bibitem{Kaka}S. Kaka, M. R. Pufall, W. H. Rippard, T. J. Silva,
S. E. Russek, and J. A. Katine, Nature (London) \textbf{437}, 389
(2005) .

\bibitem{Mancoff}F. B. Mancoff, N. D. Rizzo, B. N. Engel, and S.
Tehrani, Nature (London) \textbf{437}, 393 (2005) .

\bibitem{Grollier}J. Grollier, V. Cros, and A. Fert, Phys. Rev. B
\textbf{73}, 060409(R) (2006).

\bibitem{Slavin}A. N. Slavin and V. S. Tiberkevich, Phys. Rev. B
\textbf{72}, 092407 (2005).

\bibitem{Lee} K. -J. Lee, A. Deac, O. Redon, J. -P. Nozi\`{e}res,
and B. Dieny, Nature Materials (London) \textbf{3}, 877 (2004).

\bibitem{Berkov}D. V. Berkov and N. L. Gorn, Phys. Rev. B \textbf{71},
052403(2005); \textbf{72}, 094401(2005).

\bibitem{Alligood}K. T. Alligood, T. D. Sauer, and J. A. Yorke, \textit{Chaos:
An Introduction to Dynamical Systems} (Springer, New York, 1997).

\bibitem{Ott}E. Ott, \textit{Chaos in Dynamical Systems} (Cambridge
University Press, New York, 2002).

\bibitem{footnotes} Experimentally, the observation of the precessional
states and synchronization was made for the magnetic field at an angle
to the film plane. The primary reason is that the limit cycle or the
precessional state generally occurs at a much wider range of parameters
for the field applied at an angle to the plane than for the field
in the plane. Thus it is easier for experiments to observe the limit
cycle in the former case. For our theoretical study of chaotic dynamics,
it is natural to minimize the additional parameter (the direction
of the field) by choosing the field along the easy axis. We expect
that there is no fundamental difference with respect to the direction
of the field.

\bibitem{Serpicoa}C. Serpico, M. d'Aquino, G. Bertotti, and I. D.
Mayergoyz, J. Magn. Magn. Mater. \textbf{290-291}, 502(2005).

\bibitem{Li_Chaos}Z. Li, Y. C. Li, and S. Zhang, Phys. Rev. B \textbf{74},
054417 (2006).

\bibitem{Gonzalez-Miranda}J. M. Gonz\'{a}lez-Miranda, \textit{Synchronization
and Control of Chaos: An Introduction for Scientists and Engineers}
(Imperial College Press, London, 2004).


%\bibitem{Iaxis}One may choose the initial points in a small sphere.
%With the evolution of the dynamics, the sphere becomes an ellipsoid.
%The $i$th axis corresponds to the $i$th axis of the ellipsoid. See
%Ref.~\cite{Alligood,Gonzalez-Miranda} for more general explanation.


\bibitem{Feigenbaum}M. J. Feigenbaum, J. Stat. Phys \textbf{19},
25(1978).

\bibitem{VanDerWeele}J. P. Van Der Weele, H. W. Capel and R. Kluiving,
Physica \textbf{145A}, 425 (1987).

\bibitem{Crutchfield}J. P. Crutchfield, J. D. Farmer and B. A. Huberman,
Phys. Rep. \textbf{92}, 45 (1982)
\end{thebibliography}
\end{document}